\documentclass[%
 reprint,
 amsmath,amssymb,
 aps,
]{revtex4-2}

\usepackage{graphicx}
\usepackage{dcolumn}
\usepackage{bm}

\begin{document}

\preprint{APS/123-QED}

\title{Phase of Kerr-based few-cycle parametric amplification}

\author{Nathan Drouillard}%
\author{TJ Hammond}
 \email{tj.hammond@uwindsor.ca}
\affiliation{%
 Department of Physics\\
 University of Windsor\\
 Windsor ON N9B 3P4 Canada\\
}%

\date{\today}

\begin{abstract}
Kerr instability amplification can amplify over an octave of spectrum, a broad bandwdith supporting few-cycle pulses. However, dispersion management in this regime is crucial to maintain the ultrashort pulse duration. In our simulations, we find that the dispersion of Kerr instability amplification is near zero at the pump wavelength, and can be compensated pre-amplification to generate near-transform-limited amplified few-cycle pulses. We also find the phase of the amplified pulse depends on the seed phase and is independent of the pump, and does not significantly depend on the pump intensity. We discuss chirping the seed pulse to avoid saturation, a route for generating sub-mJ few-cycle pulses from the Kerr nonlinearity.
\end{abstract}

\maketitle


\section{Introduction}

Kerr instability amplification (KIA) has demonstrated the amplification of supercontinuum spectra spanning nearly an octave in bandwidth \cite{VampaScience2018, Ghosh2023A, Ghosh2023B}. Such broadband spectra support few-cycle pulses, where the peak field strength depends on both the dispersion and the carrier envelope phase (CEP) \cite{KrauszPhysScr2016, IshiiNatComm2014, HammondNatPhot2017}. The broad amplified spectrum drives intense ultrafast experiments such as high harmonic generation, strong-field physics, and attosecond (1~as = $10^{-18}$~s) science \cite{CorkumNatPhys2007, CalegariJPB2016, CiappinaRPP2017}.  Femtosecond (1 fs = 10$^{-15}$~s) amplifiers must amplify the broad spectrum without introducing spurious temporal and spatial phase effects that would compromise the pulse coherence.

Chirped pulse amplification (CPA) enables large gain over a broad bandwidth by temporally dispersing the seed pulse, stretching it to decrease the peak intensity and avoid multiphoton absorption and optical damage in the gain medium \cite{StricklandOptComm1985, ChenOptExp2023}. The amplified pulse is then subsequently recompressed to undo the initial dispersion, leading to intense fs pulses. However, the limited gain medium bandwidth leads to gain narrowing with each pass in the amplifier \cite{LeBlancOC1996, DurfeeIEEE1998, JiOptComm2021}. 

Optical parametric amplifiers (OPAs) can side-step this gain narrowing by several means. A noncollinear geometry (NOPA) can substantially increase the gain bandwidth, leading to amplified spectra that support few-cycle pulses \cite{CerulloRSI2003}. Multiple amplification stages can be combined with other nonlinear processes to create intense few-cycle pulses with central wavelengths from the visible to the mid-infrared \cite{RothhardtLPR2017, EluNatPhot2021, DubietisOEA2023}. Fourier-domain amplification (FOPA) tailors the gain of each spectral component, allowing for spectral shaping and tuning \cite{SchmidtNatComm2014, DallaBarbaOptExp2023}. However, these OPAs require crystals with large second-order nonlinearity, limiting the choice of available gain media. 

Conversely, KIA relies on the third-order nonlinearity, which is present in all materials, and can amplify supercontinuum spectra that support few-cycle pulses in a single step without additional pulse shaping. Furthermore, the parametric amplification process automatically allows for excellent pulse contrast without requiring pulse picking, making it a suitable seed for petaWatt scale few-cycle sources \cite{WangOL2019}. We propose that since KIA has a larger gain and bandwidth than second-order OPAs, it could be used as a first-stage parametric amplifier for few-cycle pulses. However, several phase effects must be studied to ensure that the phase is maintained without introducing significant pulse front distortion. 

In this work, we simulate the KIA amplification process to study the effects of amplification on the phase. We investigate the dispersion in the amplification process and how controlling the seed phase allows for amplified few-cycle pulses near the transform limit. We also discuss the effect of the seed and pump CEP on the amplified pulse, as well as pump intensity jitter. Although saturation distorts the amplified pulse, we can avoid this effect by pre-chirping the pulse to decrease the seed peak intensity, enabling Kerr instability chirped-pulse amplification (KICPA).

\section{Amplified pulse dispersion}

As with other parametric processes, KIA requires satisfying the phase-matching condition to maximize the gain. Because of the frequency-dependent refractive index of the gain medium, the phase-matching of this third-order process is not satisfied collinearly in general, and requires a non-collinear geometry. This (internal) angle is given by \cite{Ghosh2023B}
\begin{equation}
\cos\theta_s = \frac{4 (n_p^2 - n_2 I_p) \omega_p^2 + n_s^2 \omega_s^2 - n_i^2 \omega_i^2}{4 n_s \omega_s \omega_p \sqrt{n_p^2 - n_2 I_p}}, \label{eq:phasecurve}
\end{equation}
where $n_{p,s,i} = n(\omega_{p,s,i})$ are the indices of refraction of the pump, signal, and idler, respectively, and $n_2$ is the nonlinear index and $I_p$ is the peak intensity.

Amplifying supercontinuum spectra then requires careful spatial dispersion compensation, limiting the applicability of this scheme. However, when pumped at high intensity, MgO has shown to have a nearly constant angle that maximizes the amplification over an octave of bandwidth \cite{Ghosh2023A, Ghosh2023B}. We simulate the amplification of few-cycle pulses at a constant angle of $4.5^\circ$ (external angle) in 0.5~mm MgO when pumped at 1064~nm with peak intensity of $1.5\times10^{17}$ W/m$^2$, unless otherwise specified.

Our simulations are performed in two dimensions to account for the transverse phase matching criterion. We solve the forward Maxwell equation (FME) \cite{HousakouPRL2001, BergeRPP2007} by fourth order Runge-Kutta (RK4), fast Fourier transforming position to momentum spaces for wavefront propagation \cite{Ghosh2023A, Ghosh2023B}. We have previously found consistent agreement between KIA theory and our simulations for amplitude and angle dependence. We now use these simulations to investigate the phase dependence of KIA.

\begin{figure}[ht]
\includegraphics[width=1\columnwidth]{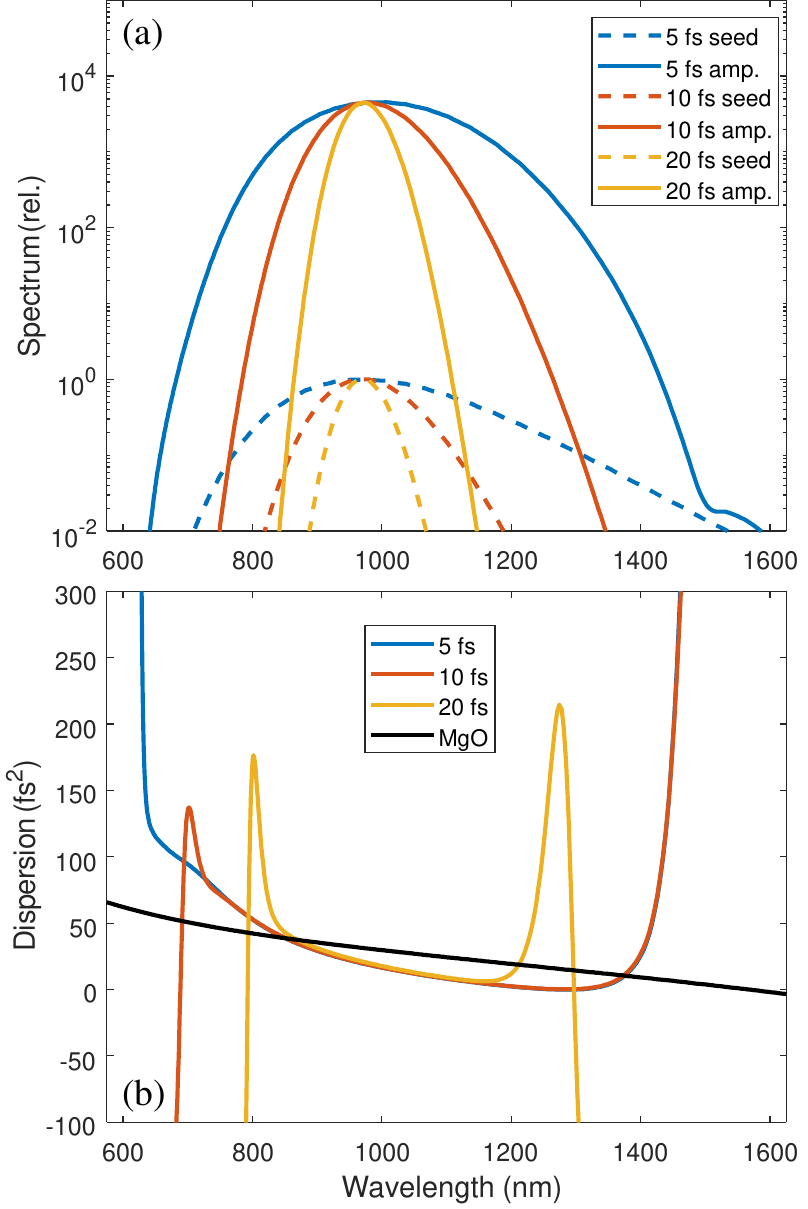}
\caption{Amplification and GDD of 5, 10, and 20 fs pulses centered at 970$~$nm from 1064$~$nm pump; non-collinear angle is 4.5$^\circ$ external angle. (a) The pulse durations are 6.2$~$fs (8.9$~$fs), 10.4$~$fs (11.3$~$fs), and 20.0$~$fs (20.2$~$fs) for the 5, 10, and 20$~$fs amplified pulse cases for transform limit (including GDD). (b) The dispersion of the amplified pulses follows a similar spectral dependence over the amplified portion of the spectrum, and is lower than the material dispersion imparted by 0.5$~$mm of MgO over a substantial portion of the bandwidth.}
\centering
\end{figure}

In Fig. 1, we simulate amplifying ultrashort seed pulses centered at 970$~$nm to understand the effect of seed bandwidth on the amplified spectrum and phase. For the amplified spectra in (a), we find that there was negligible change for spectra supporting pulses as short as 10~fs full width at half maximum (FWHM) in duration by comparing the dashed (seed) to amplified (solid) spectra. Gain narrowing for few-cycle pulses increases the pulse duration from 5~fs to 6.2~fs in the transform limit (TL). In all three cases, the relative amplification factor is $4400\times$ at 970~nm, or a gain of $g = 16.8$/mm.

While KIA predicts zero dispersion at the pump wavelength \cite{NesrallahOptica2018}, the initial $\sim 150$~\textmu m does not lead to the seed (signal) amplification, but rather idler creation \cite{Ghosh2023B}. Once the idler reaches the amplitude of the seed, the dispersion goes to zero. Thus, the dispersion near the pump is 10.1 fs$^2$, significantly less than the material dispersion of 26.4~fs$^2$. However, higher order material dispersion terms still contribute, as shown in Fig. 1(b), and we can estimate the dispersion of the amplified pulse near the peak of the spectrum by the third order dispersion (TOD) of the gain medium. Away from the spectrum peak, the dispersion also depends on the pulse bandwidth. For the 20~fs case, because the amplified bandwdidth is well within the region of maximum amplification, the dispersion follows the TOD of MgO. As the seed bandwidth increases to the amplification bandwidth limit (the 5~fs case), the dispersion deviates from the MgO TOD and becomes spectral-shape dependent. 

\begin{figure}[ht]
\includegraphics[width=1\columnwidth]{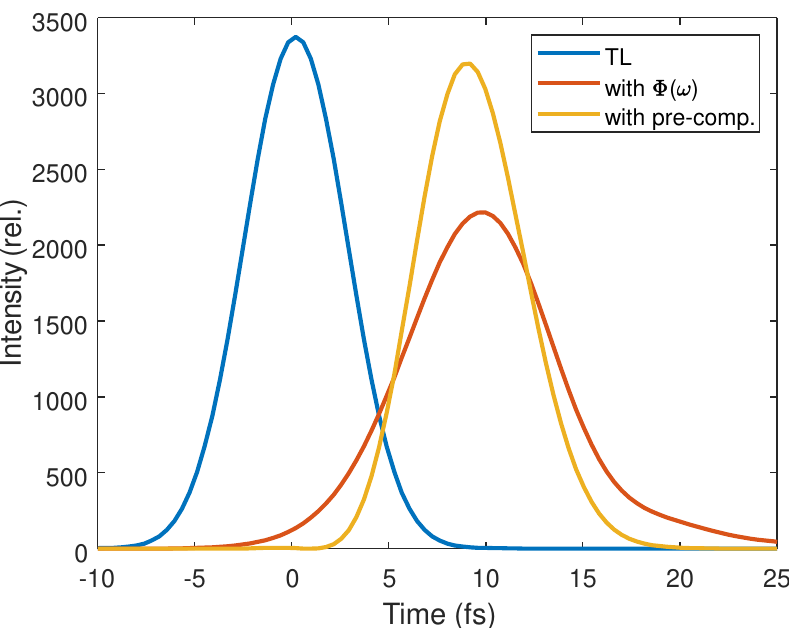}
\caption{A 5~fs 970~nm seed pulse amplified to transform limited duration (blue) 6.2~fs, accounting for phase (orange) becomes 8.9~fs. By pre-compensating for the GDD incurred by the amplification process (yellow), the pulse duration is 6.5~fs. A 6th order polynomial fits the dispersion of the amplified pulse, and its negative is applied to the seed pulse to pre-compensate for the dispersion.}
\centering
\end{figure}

Because the dispersion is smooth over the majority of the amplified pulse bandwidth, we can compensate for it to amplify compressed pulses. Although it is common in laser amplifiers to compensate for the dispersion post-amplification, such as with CPA, we demonstrate here in our simulations that it is possible to compensate pre-amplification, maximing the useable peak power. As shown in Fig. 2, the transform limited (TL) amplified pulse increases in duration to 6.2~fs (from a 5~fs seed). Including the amplification dispersion, the pulse duration becomes 8.9~fs. We can compensate for the amplified pulse phase by shaping the seed, fitting its phase to a 6th order polynomial (accounting for GDD and higher orders) \cite{TournoisOC1997, WeinerOptComm2011, LoriotOE2013}. This amplified pulse duration is 6.5~fs, or two-cycle at 970~nm central wavelength.

\section{Carrier envelope phase}

In spite of the significant dispersion imparted by the stretching and compression steps of chirped-pulse amplification in laser amplifiers, the carrier envelope phase (CEP) of the seed oscillator is maintained through to the amplified pulse. Although the amplification process itself can increase in the CEP jitter to 100~mrad \cite{HongAPL2006, FrankRSI2012, LuckingOL2014}, drifts can be compensated by stabilizing the oscillator CEP, pump intensity control, and other feedback (and feedforward) mechanisms \cite{BalciunasOL2011, BalciunasOL2014, CunninghamAPL2015}.

\begin{figure}[ht]
\includegraphics[width=1\columnwidth]{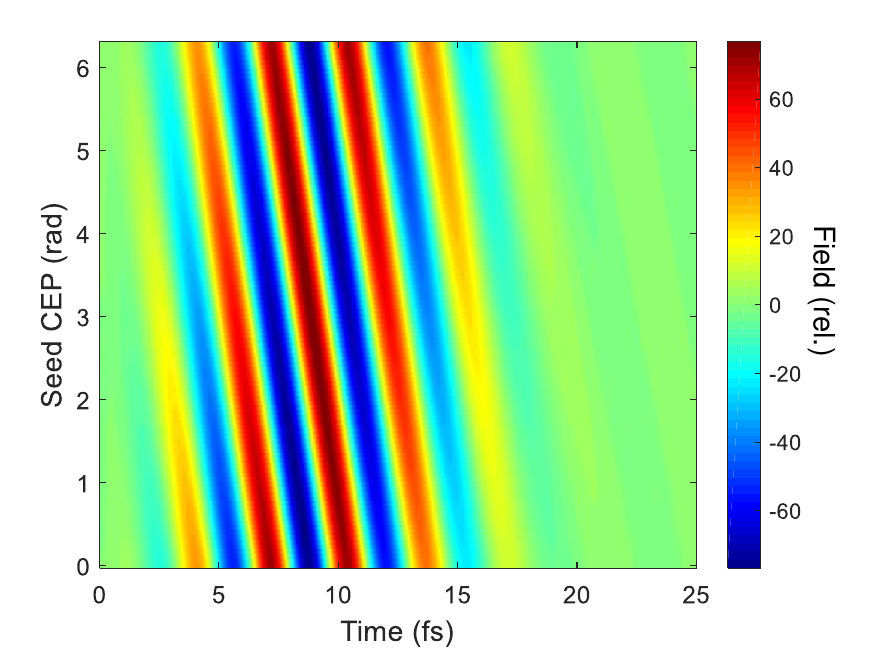}
\caption{Varying the seed CEP changed the amplified field phase, but varying the pump CEP did not. The field is plotted relative to the maximum seed field amplitude.}
\centering
\end{figure}

In our simulations, we found a similar effect in that the CEP of the oscillator was directly related to the CEP of the amplified pulse, as shown in Fig. 3. We vary the seed CEP over $2\pi$ and find that there is a direct relationship to the amplified field phase. We note that the amplified field phase is independent of the pump CEP. Because the ideal pump would be significantly longer in duration than the seed, the pump CEP independence benefits the simplicity of the amplification scheme because of the difficulty in stabilizing narrow bandwidth pulses.

\begin{figure}[ht]
\includegraphics[width=8.6cm]{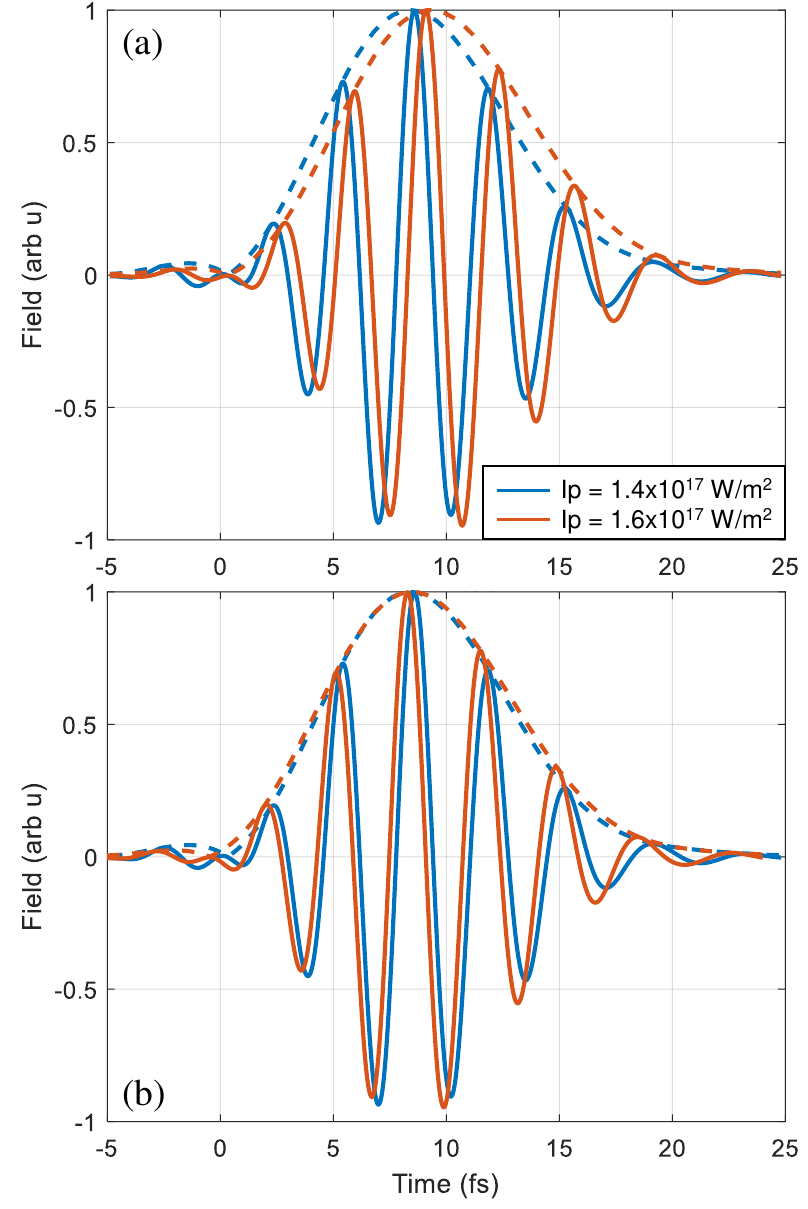}
\caption{Effect of pump intensity on amplified pulse phase; the seed dispersion is pre-compensated. (a) There is a 780~as delay in the envelope as the pump intensity increases from 1.4 to $1.6\times10^{17}$~W/m$^2$, or 55~as/\% pump intensity change. (b) After correcting the group delay, we also measure a CEP shift of 0.54~rad over this intensity range, or 38~mrad/\% pump intensity change.}
\centering
\end{figure}

The amplified pulse CEP depends on the pump intensity, however. As the pump intensity increases from $1.4\times10^{17}$~W/m$^2$ to $1.6\times10^{17}$~W/m$^2$, there is an increased group delay of 780~as, as shown in Fig. 4(a). This group delay agrees with the expected delay due to the nonlinear index 
\begin{align}
    T_d = \frac{L}{c}\frac{n_2}{n(\omega_p)}\Delta I_p
\end{align}
where $\Delta I_p$ is the pump intensity jitter. Compared to the pump intensity of $1.5\times10^{17}$~W/m$^2$, this corresponds to a 55~as/\% pump intensity change. This pump intensity to delay jitter could lead to difficulty in optical heterodyne detection experiments \cite{JonesTAP2004}. 

After compensating for this jitter, there is also a CEP change of 0.54~rad over this intensity range, or 38~mrad/\% pump intensity change. For CEP dependent experiments, KIA will benefit from pump lasers with well-controlled pulse power. The CEP jitter stems from the intensity dependent phase matching criterion. Because the phase matching angle depends on the pump intensity, the emerging field also has a slight angle dependence, resulting in a shift in the CEP. Although we found this CEP change to be consistent over these parameters, we also find that the CEP change depends on the pump and seed central wavelengths, the bandwidth, and the relative angles.

\section{Kerr instability chirped pulse amplification}

\begin{figure}[h]
\includegraphics[width=1\columnwidth]{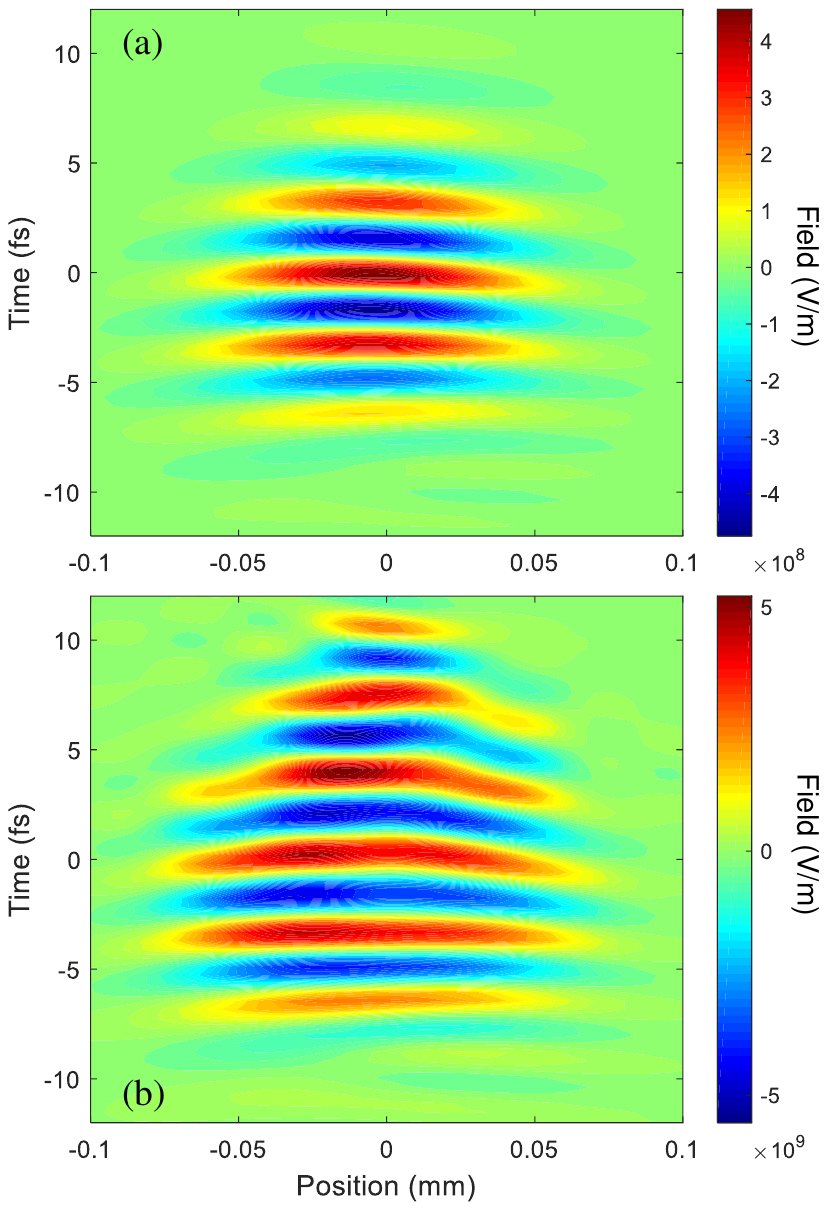}
\caption{Saturation effects on the spatio-temporal field evolution. (a) Relative seed intensity $I_s/I_p = 10^{-6}$ leads to a compressed amplified pulse without spatial chirp. (b) When $I_s/I_p = 6.25\times10^{-4}$, as the amplified pulse intensity reaches $>1\%$ of the pump (amplification factor $\approx10^4$), the amplification is saturated and the field becomes distorted leading to walkoff and increased duration.}
\centering
\end{figure}

Once the seed reaches $>1\%$ of the pump, the seed and idler saturate, leading to a cascaded four-wave mixing process that generates higher-order beamlets \cite{WeigandApplSci2015, JACOL2021}. Although these beamlets can be recombined to generate a few-cycle pulse, the difficulty in compensating for the angular dispersion limits this technique for attosecond science experiments \cite{KobayashiJPB2012, HeApplSci2014, WeigandApplSci2015}. At saturation, we find that the amplified field becomes distorted, leading to beam walk-off and pulse front tilt. Such a result is comparable to noncollinear optical parametric chirped pulse amplifiers \cite{GireeOptExp2017}.

\begin{figure}[h]
\includegraphics[width=1\columnwidth]{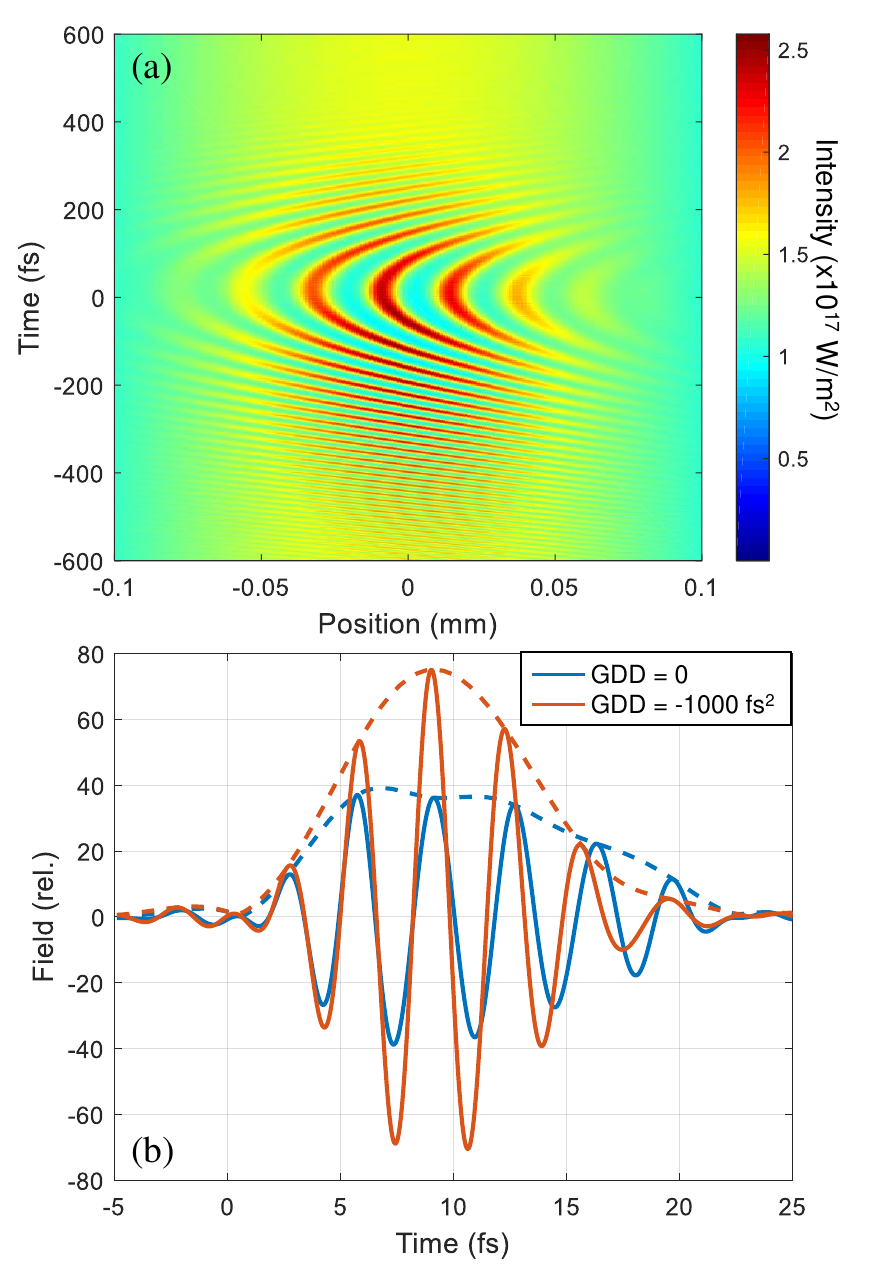}
\caption{Applying GDD $= -1000$~fs$^2$ to a 5~fs pulse decreases the peak intensity by $\sim 100\times$, avoiding the saturation effects. (a) Space-time intensity profile of pump and amplified pulse shows a parabolic interference pattern signifying the initial GDD. (b) Amplified field relative to seed for unchirped (blue) and chirped (orange) seed fields.}
\centering
\end{figure}

We show the effect of saturation in Fig. 5. When the relative seed to pump intensity, $I_s/I_p = 10^{-6}$ and the seed phase is pre-compensated, the amplified pulse is below saturation, as shown in (a). In this case, the field demonstrates no spatial distortion to the wavefronts, and the pulse duration is 6.5$~$fs, close to the input pulse duration of 5~fs. Conversely, with a relative intensity of $6.25\times10^{-4}$, the saturated amplified pulse is spatially and temporally chirped, as shown in (b).

We find we can avoid saturation by decreasing the peak intensity by chirping the seed. As shown in Fig. 6, we simulate the amplification of a 5~fs seed pulse stretched with GDD $= -1000$~fs$^2$ to decrease its peak intensity by a factor of 110. With chirp, the peak seed intensity relative to the pump is $I_s/I_p = 5.7\times10^{-6}$. In this case, the pump, with a central wavelength of 1064~nm, has a top-hat temporal profile of 1500~fs duration. The total intensity of pump and seed is shown in (a) at the exit of the MgO. The parabolic interference demonstrates the seed dispersion, and is below saturation despite the significant intensity modulation. After amplification, we impose +1000~fs$^2$ to recompress the pulse, as shown in (b). Without dispersion control, the pulse saturates and the amplified field is $\sim 40\times$ greater than the seed and the pulse duration increases (blue). With the dispersion control, we avoid saturation and the pulse is nearly transform limited (orange, pulse duration 6.5~fs) and the peak field strength is $\sim 75\times$ the seed.

\section{Conclusions}

We find that the GDD of KIA is near zero at the pump wavelength, and follows the TOD over a significant portion of the amplified pulse bandwidth. This imparted phase from the amplification process can be compensated by controlling the seed phase to amplify nearly transform limited few-cycle pulses. The amplified pulse CEP depends on the seed and is nearly independent of the pump. This scheme offers a simple route to generating intense CEP-stable few-cycle pulses. Saturation significantly distorts the amplified field, leading both spatial and temporal chirp. However, chirping the seed can avoid saturation, increasing the amplified pulse peak intensity.

\begin{acknowledgments}
We acknowledge funding from Natural Sciences and Engineering Research Council of Canada (RGPIN-2019-06877) and the University of Windsor Xcellerate grant (5218522). TJH thanks Thomas Brabec and Claire Duncan for useful conversations.
\end{acknowledgments}


\bibliography{biblio}

\end{document}